\documentclass[preprint,12pt]{elsarticle}

\usepackage{multicol}
\usepackage{wrapfig} 
\usepackage{flushend} 
\usepackage{amsmath}
\usepackage{booktabs}
\usepackage{graphicx}
\usepackage{tabularx}
\usepackage{makecell} 
\usepackage{multirow}
\usepackage{array}
\usepackage{CJKutf8}
\usepackage{amssymb}
\usepackage{amsmath}
\usepackage{float}

\journal{arxiv}

\begin{document}

\begin{frontmatter}



\title{TCKAN: A Novel Integrated Network Model for Predicting Mortality Risk in Sepsis Patients}


\author[label1]{Fanglin Dong} 
\author[label2]{Shibo Li}
\author[]{Weihua Li\corref{cor1}}

\affiliation[label1]{organization={Yunnan University},
            addressline={dongfanglin@stu.ynu.edu.cn}, 
            city={Kunming},
            postcode={650000}, 
            state={Yunnan Province},
            country={China}}
\affiliation[label2]{organization={Yunnan University},
            addressline={lishibo@stu.ynu.edu.cn}, 
            city={Kunming},
            postcode={650000}, 
            state={Yunnan Province},
            country={China}}
\cortext[cor1]{Corresponding author: liweihua@ynu.edu.cn}

\begin{abstract}
Sepsis poses a major global health threat, accounting for millions of deaths annually and significant economic costs. Accurately predicting the risk of mortality in sepsis patients enables early identification, promotes the efficient allocation of medical resources, and facilitates timely interventions, thereby improving patient outcomes. Current methods typically utilize only one type of data—either constant, temporal, or ICD codes. This study introduces a novel approach, the Time-Constant Kolmogorov-Arnold Network (TCKAN), which uniquely integrates temporal data, constant data, and ICD codes within a single predictive model. Unlike existing methods that typically rely on one type of data, TCKAN leverages a multi-modal data integration strategy, resulting in superior predictive accuracy and robustness in identifying high-risk sepsis patients. Validated against the MIMIC-III and MIMIC-IV datasets, TCKAN surpasses existing machine learning and deep learning methods in accuracy, sensitivity, and specificity. Notably, TCKAN achieved AUCs of 87.76\% and 88.07\%, demonstrating superior capability in identifying high-risk patients. Additionally, TCKAN effectively combats the prevalent issue of data imbalance in clinical settings, improving the detection of patients at elevated risk of mortality and facilitating timely interventions. These results confirm the model’s effectiveness and its potential to transform patient management and treatment optimization in clinical practice. Although the TCKAN model has already incorporated temporal, constant, and ICD code data, future research could include more diverse medical data types, such as imaging and laboratory test results, to achieve a more comprehensive data integration and further improve predictive accuracy.
\end{abstract}



\begin{keyword}
Sepsis Patient \sep Mortality Risk Prediction\sep Kolmogorov-Arnold Network 



\end{keyword}

\end{frontmatter}


\section{Introduction}
Sepsis, a life-threatening condition characterized by a systemic inflammatory response to infection\cite{1}, poses a significant global health burden\cite{3,4}. Accurate prediction of sepsis mortality risk is essential for efficient resource allocation and improved patient outcomes \cite{7,chen2022towards,8}. 

To effectively allocate resources and improve patient outcomes, numerous models have been developed to predict the mortality risk associated with sepsis. Scoring tools \cite{11,13} are widely used to assess the severity of sepsis. While various scoring tools have been developed, these tools are predominantly based on specific populations and clinical settings from their time of development, leading to reduced predictive accuracy as demographics and practices evolve. Additionally, these methods rely on a limited set of static physiological parameters that do not fully capture the dynamic changes in patient conditions\cite{14,15}. 

The rise of Electronic Health Records (EHRs) has enhanced the availability of medical data, crucial for developing predictive models using physiological, laboratory, and demographic information \cite{9}. This has spurred the application of machine learning in sepsis mortality prediction. For instance, Taylor et al. demonstrated that a Random Forest model, utilizing over 500 clinical variables, outperformed Logistic Regression in predicting sepsis mortality \cite{17, 20}. Similarly, Zhang et al. developed a LASSO-based score using the MIMIC-III dataset \cite{18}, though external validation is still required. Kong et al. extended this work by developing four models—LASSO \cite{19}, Random Forest \cite{20}, Gradient Boosting Machine \cite{21}, and Logistic Regression—based on 86 features from MIMIC-III to predict in-hospital mortality \cite{22}. However, traditional machine learning methods often encounter challenges when dealing with large feature sets and time-series data, which can lead to overfitting and reduced effectiveness in sepsis mortality prediction.

Deep learning, a pivotal branch of machine learning, has recently made significant strides in biomedicine, with applications in genomics \cite{li2023learning, guo2022context}, metabolomics \cite{xue2024rt}, and proteomics \cite{jia2024metafluad}. In health informatics, particularly in mortality prediction, deep learning has demonstrated its capability to automatically extract features from large datasets and handle time-series data, thereby enhancing predictive accuracy \cite{ghoshroy2023ai}. For instance, For example, Su et al. used a two-layer MLP model to predict 30-day mortality in sepsis patients, but the lack of temporal data limited its predictive power \cite{28}. Cheng et al. utilized Convolutional Neural Network (CNN)\cite{29}, Long Short-Term Memory (LSTM) \cite{30}, and Random Forest (RF) to analyze sepsis patients' temporal vital signs data, predicting in-hospital mortality\cite{31}. However, their study was confined to a minimal set of seven features, which might have overlooked more complex variables influencing patient outcomes. Similarly, Gong et al. developed a Temporal Convolutional Networks (TCN)-based model that integrated time-series data with four vital signs to predict near-term mortality risks in sepsis patients, yet it also limited its analysis to a few variables\cite{32}. Moreover, Recent studies support the benefits of integrating multiple data sources to improve predictive model performance \cite{roy2022artificial}. and knowledge-guided approaches can improve diagnostic predictions \cite{li2023knowledge}. However, existing studies often rely on limited sets of features, fail to effectively integrate temporal and constant data, or overlook the incorporation of domain-specific additional knowledge.

To tackle the above issues, this study introduces the Temporal-Constant Kolmogorov\allowbreak-Arnold Network (TCKAN) to enhance sepsis mortality prediction. Building upon the GRU-D\cite{33} and KAN architectures\cite{34}, TCKAN effectively integrates temporal data, constant data, and diagnostic ICD codes into a unified model. Figure~\ref{fig:figure 1} illustrates the main flowchart of the TCKAN model. Our principal contributions are outlined as follows:

\begin{figure}[htb]
    \centering
    \includegraphics[width=1\textwidth, page=1]{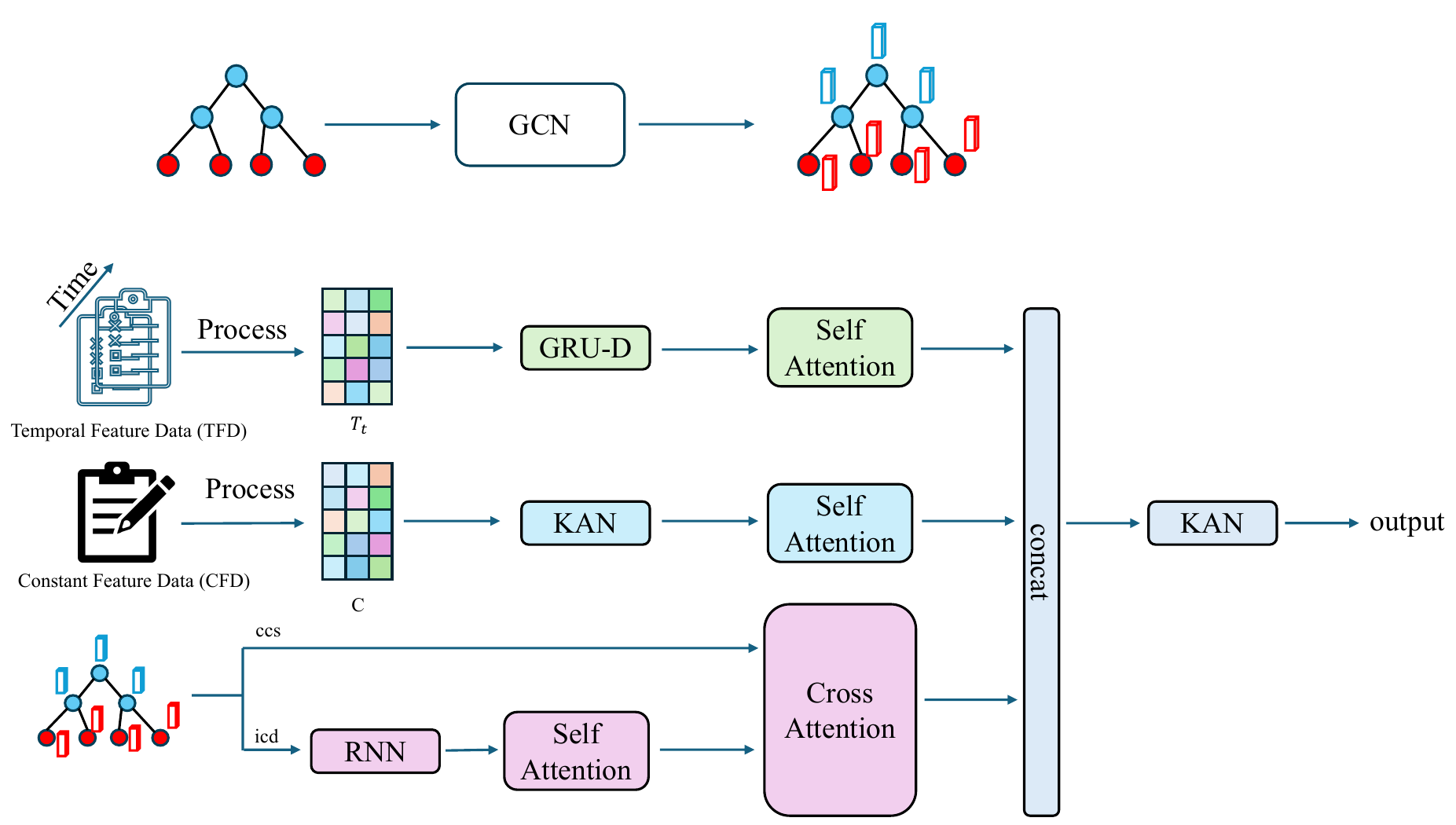}
    \caption{Main flowchart of the TCKAN}
    \label{fig:figure 1}
\end{figure}

\begin{enumerate}[(1)]
    \item Employing graph neural networks to integrate diagnostic ICD codes and Clinical Classification Software (CCS) for a more informative representation of patient characteristics.
    \item Developing an effective fusion mechanism using GRU-D and KAN to capture the dynamic progression of patient states.
    \item Demonstrating TCKAN's superior prediction accuracy and robustness over existing methods compared to existing methods, validated across two datasets. 
\end{enumerate}

\section{Methods}
\subsection{Datasets}
This study utilizes the MIMIC-III\cite{18} and MIMIC-IV\cite{35} datasets, both derived from the Intensive Care Units (ICU) at Beth Israel Deaconess Medical Center, USA. These datasets were collaboratively created by the Massachusetts Institute of Technology's Computer Science and Artificial Intelligence Laboratory (MIT CSAIL) and the medical center. MIMIC-III, publicly available, encompasses detailed health records of approximately 60,000 patients from 2001 to 2012, encompassing patient demographics, vital signs, laboratory test results, medication records, nursing notes, radiology reports, and discharge summaries. This database supports clinical research and the development of machine learning and artificial intelligence algorithms. MIMI\-C-IV, an enhancement of MIMIC-III, includes ICU data from 2012 to 2019, featuring improved data quality and expanded clinical data types over a longer record span. These enhancements facilitate complex medical research and sophisticated data analysis. This study leverages these rich datasets to deeply analyze diagnostic information, temporal changes in vital signs, and other pertinent clinical features of sepsis patients. Authorization to use the MIMIC databases was obtained, and data extraction was conducted. Ethical approval was granted by the Institutional Review Board of Beth Israel Deaconess Medical Center. Compliance with HIPAA regulations ensured de-identification of all Protected Health Information (PHI), eliminating the need for patient consent and maintaining strict ethical and privacy standards to protect patient privacy effectively.

\subsection{Data Preprocessing}
\begin{figure}[htb]
    \centering
    \includegraphics[width=1\textwidth, page=1]{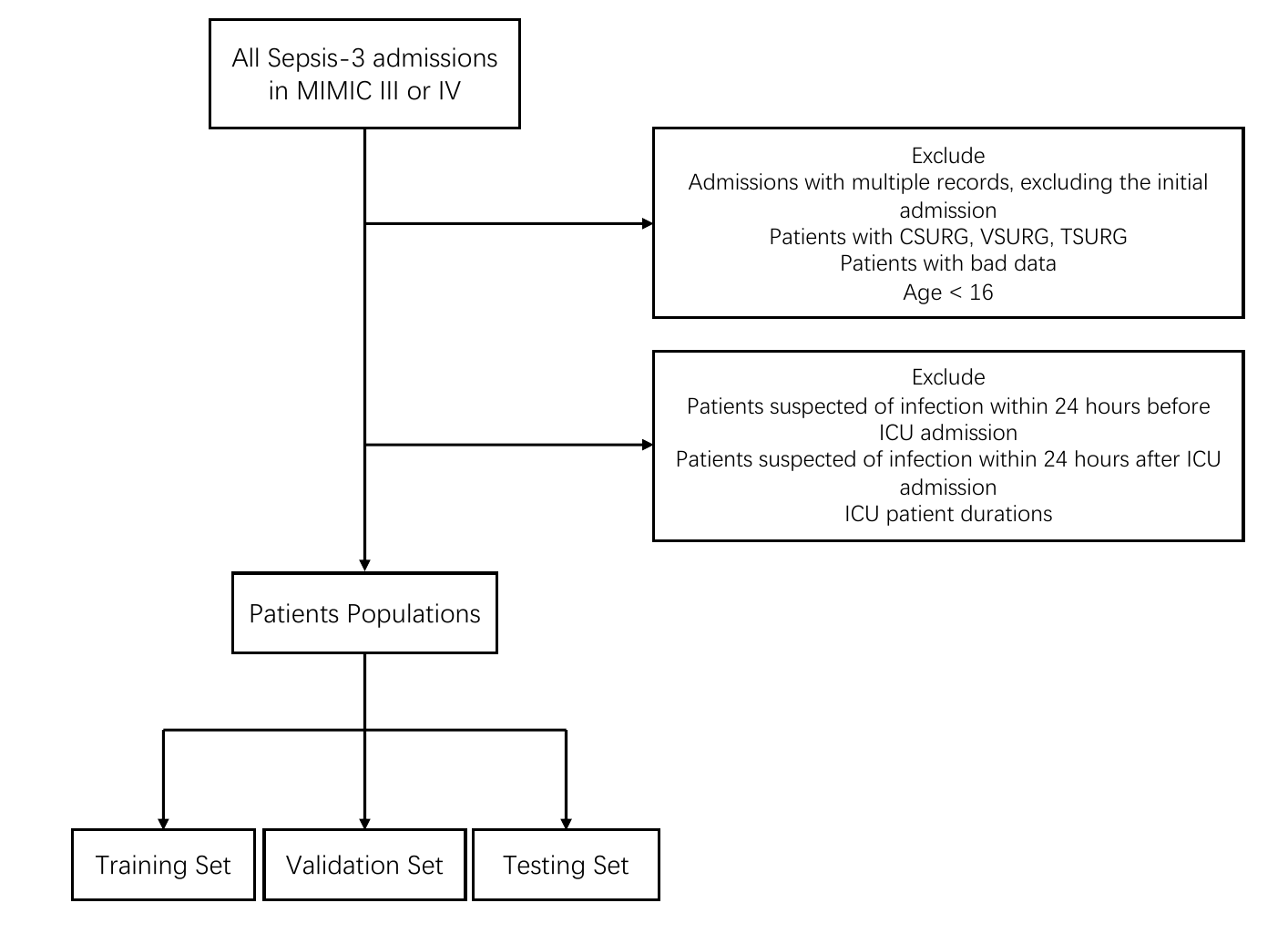}
    \caption{Process for extracting data from MIMIC-III and MIMIC-IV}
    \label{fig:figure 2}
\end{figure}

Based on the Sepsis-3 definition\cite{36}, we selected sepsis patients from the MIMIC-III and MIMIC-IV datasets who had a SOFA score of 2 or higher, as illustrated in Figure~\ref{fig:figure 2}. Initially, we excluded patients under the age of 16 and those with corrupted data. We also eliminated patients with suspected infections occurring more than 24 hours before or after ICU admission, due to the ambiguous timing of their infection diagnoses, which could compromise the study's accuracy. Furthermore, we excluded patients with ICU stays shorter than 24 hours, as their incomplete clinical data insufficiently support the model. Additionally, patients who underwent cardiac, thoracic, or vascular surgery were omitted because their unique surgical treatments and care requirements could adversely affect the model's generalizability. For sepsis patients with multiple hospital admissions, we used data from their initial admission. We extracted a total of 10,567 patients from the MIMIC-III dataset (1,190 positive cases and 9,377 negative cases) and 11,540 patients from the MIMIC-IV dataset (1,274 positive cases and 10,266 negative cases).

Using these criteria, we extracted patients' constant and temporal data, along with diagnostic ICD coding information, from the MIMIC-III and MIMIC-IV databases. Constant data includes basic demographic information like gender, age, and weight, which provides essential background on the patients. Temporal data encompasses detailed physiological signs and laboratory test results recorded during the ICU stay, such as temperature, heart rate, blood pressure, respiratory rate, hemoglobin levels, and white blood cell count, reflecting the immediate health status and condition changes of the patients. Additionally, diagnostic ICD codes were transformed into vector representations via graph embedding methods, allowing for a comprehensive integration of patients' constant characteristics, temporal features, and diagnostic information. The patient features extracted from the MIMIC-III and MIMIC-IV databases were obtained using SQL queries. Table \ref{tab:Table 1} provides a detailed listing of the features that were selected and extracted using SQL methods from the temporal and constant data available in the MIMIC-III and MIMIC-IV databases.

\begin{table}[h]
\centering
\caption{Overview of Patient Features Extracted from MIMIC-III and MIMIC-IV Databases}
\label{tab:Table 1}
\begin{tabularx}{\textwidth}{@{}Xp{10cm}@{}} 
\toprule
    Feature type       & \multicolumn{1}{c}{Name} \\ \midrule
Temporal Feature & \small{Platelet, MCH, MCHC, Fibrinogen, Basophils, Lymphocytes, HCO3, Troponin-t, Sodium, Urine output, Bilirubin, HR, Lactic acid, Albumin, Calcium, Glucose, PH, Creatinine, Eosinophils, Total CO2, SBP, RBC, Chloride, ALT, Fraction inspired oxygen set, Alkaline phosphate, PaO2, GCS, PT, SpO2, Anion gap, DBP, Magnesium, BUN, WBC, Glucose, Temperature, MCV, Troponin-i, Potassium, Neutrophils, Hemoglobin, PTT, Lactate, PaCO2, FiO2, Hematocrit, AST, Monocytes, Phosphate, INR, MBP, RR} \\ \midrule
Constant Feature & \small{Weight, GCS(motor,verbal,eyes), Insurance, Age, First careunit, Hosp adm time, Ethnicity, 24h urine output, Gender, Admission type} \\
\bottomrule
\end{tabularx}
\end{table}

In preprocessing the constant data extracted from patients, we employed specific strategies tailored to different types of missing data. For continuous variables, such as age and weight, missing values were imputed with zero to maintain data completeness and consistency without introducing bias, thereby ensuring that the model's training is not negatively impacted. For categorical variables, such as gender and race, one-hot encoding was utilized, converting each category into a binary vector where each category is represented by a unique dimension. This method prevents the imposition of unintended ordinal relationships among categories, thereby avoiding any misleading effects on the model's training.

\begin{figure}[htb]
    \centering
    \includegraphics[width=1\textwidth, page=1]{
    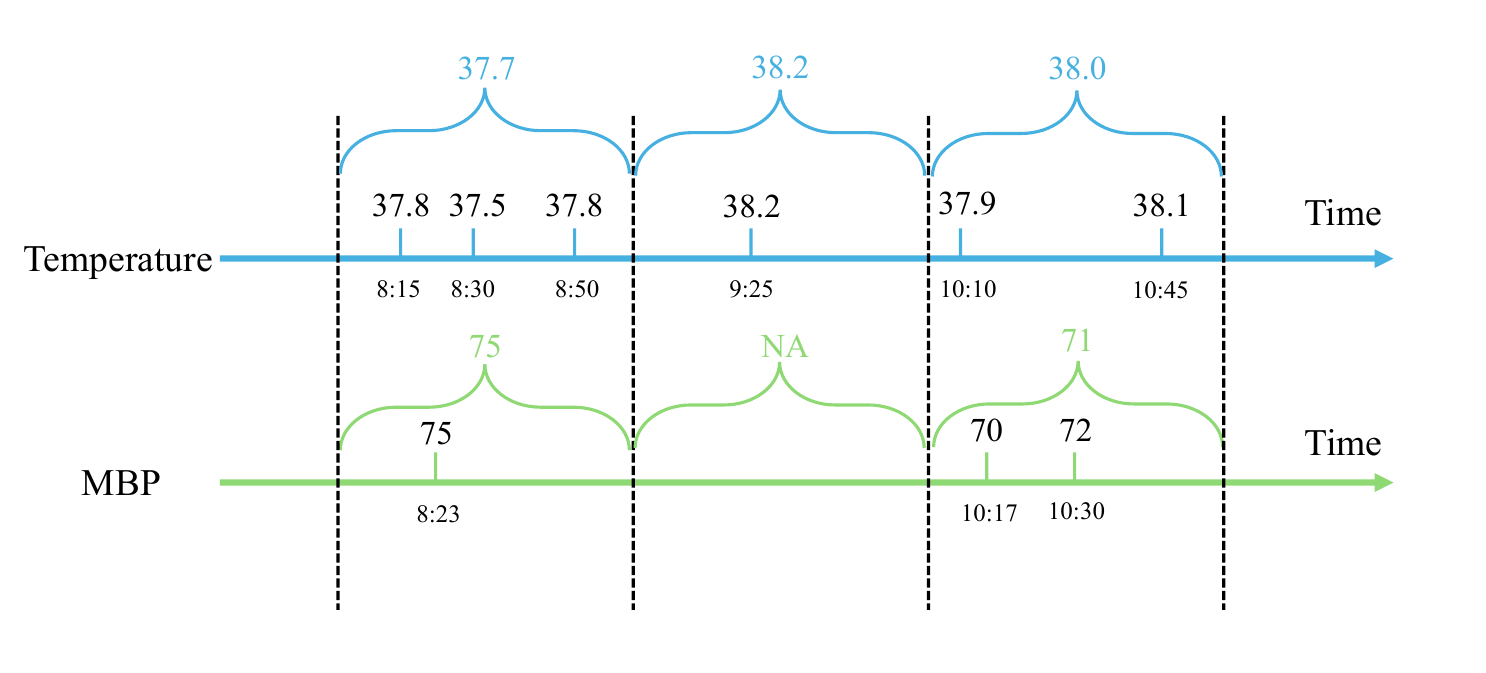}
    \caption{Sequential measurements of temperature and mean blood pressure (MBP) are recorded at specified time intervals}
    \label{fig:figure 3}
\end{figure}

For temporal data, we adhered to strict time window and interval strategies. Temporal data were extracted from the 24 hours preceding ICU admission and were sampled at 1-hour intervals. For each interval, if a single measurement was available, it was used directly; if multiple measurements were recorded, they were averaged to represent the feature level during that hour. In intervals where no measurements were recorded, values were marked as missing. After processing the constant and temporal features of the patients, we applied mean-variance normalization to the continuous variables within these features. We calculated the mean and standard deviation of each variable, using these parameters to adjust the data. This ensured that all continuous variables achieved a standardized form with zero mean and unit variance. Figure~\ref{fig:figure 3} illustrates the processing of temporal data. We will address the treatment of missing values in temporal data during subsequent model processing.

To enhance the diagnostic information of patients, we implemented the methodology of Li et al.\cite{li2023knowledge}, converting extracted patient visit ICD-9 code sequences into corresponding Clinical Classifications Software (CCS) code sequences. The CCS system groups ICD-9 codes to simplify and summarize complex diagnostic information, facilitating more efficient and intuitive analysis. Table \ref{tab:Table 2} presents data from the multi-level ICD-9-CM CCS. To construct a comprehensive diagnostic information graph, we built a medical concept graph, G=(V,E), on the MIMIC-III and MIMIC-IV datasets, following the method of Li et al. In this graph, V represents the nodes, which include ICD diagnostic codes and CCS medical ontology, while E denotes the edges that connect these nodes. This setup forms a directed tree-like structure that visually represents the hierarchical relationships between ICD codes and CCS groupings. For processing ICD and CCS codes, we applied Li et al.’s method using a two-layer graph convolutional network (GCN). Table \ref{tab:Table 3} displays details of the medical concept graph. The layers of the GCN are computed as follows:
\begin{equation*}
    H^{(l+1)} = \sigma\left(\tilde{D}^{-\frac{1}{2}} \tilde{A} \tilde{D}^{-\frac{1}{2}} H^{(l)} W^{(l)}\right)  
\end{equation*}
where $H^{(l)}$ is the node feature matrix at layer $L$, $H^{(0)}$ the initial feature representation, $\tilde{A}$ the adjacency matrix with self-connections, $\tilde{D}$ the degree matrix, $W^{(l)}$ the trainable weight matrix, and $\sigma$ the activation function. This two-layer GCN effectively embeds ICD and CCS codes into vector space.

\begin{table}[]
\caption{Example of an ICD-9-CM Hierarchy Table with CCS}
\label{tab:Table 2}
\resizebox{\columnwidth}{!}{%
\begin{tabular}{@{}ccccccc@{}}
\toprule
\begin{tabular}[c]{@{}c@{}}ICD-9-CM\\ CODE\end{tabular} & \begin{tabular}[c]{@{}c@{}}CCS\\ LVL\\ 1\end{tabular} & \begin{tabular}[c]{@{}c@{}}CCS\\ LVL\\ 1 Label\end{tabular}                   & \begin{tabular}[c]{@{}c@{}}CCS\\ LVL\\ 2\end{tabular} & \begin{tabular}[c]{@{}c@{}}CCS\\ LVL\\ 2 Label\end{tabular}                & \begin{tabular}[c]{@{}c@{}}CCS\\ LVL\\ 3\end{tabular} & \begin{tabular}[c]{@{}c@{}}CCS\\ LVL\\ 3 Label\end{tabular}                                  \\ \midrule
5761                                                    & 9                                                     & \begin{tabular}[c]{@{}c@{}}Diseases of \\ the digestive\\ system\end{tabular} & 9.7                                                   & \begin{tabular}[c]{@{}c@{}}Biliary tract\\ disease {[}149.{]}\end{tabular} & 9.7.6                                                 & \begin{tabular}[c]{@{}c@{}}Other biliary\\ tract disease\end{tabular}                        \\
V1009                                                   & 2                                                     & Neoplasms                                                                     & 2.2                                                   & \begin{tabular}[c]{@{}c@{}}Other\\ gastrointestinal\\ cancer\end{tabular}  & 2.2.5                                                 & \begin{tabular}[c]{@{}c@{}}Cancer of\\ other GI\\ organs;\\ peritoneum{[}18.{]}\end{tabular} \\ \bottomrule
\end{tabular}%
}
\end{table}

\begin{table}[h] 
\centering 
\caption{Number of CCS and ICD Nodes in the MIMIC-III and MIMIC-IV Datasets}
\label{tab:Table 3}
\begin{tabular}{ccc}
\toprule
                        & \begin{tabular}[c]{@{}c@{}}MIMIC\\ III\end{tabular} & \begin{tabular}[c]{@{}c@{}}MIMIC\\ IV\end{tabular} \\
\midrule
The number of CCS nodes & 269                                                 & 274                                                \\
The number of ICD nodes & 4616                                                & 5066                                               \\
\bottomrule
\end{tabular}
\end{table}

\subsection{The Architecture of TCKAN}
The TCKAN model architecture includes three components: (1) Temporal data processed by GRU-D (Gated Recurrent Unit with Decay) to generate hidden representations; (2) ICD and CCS codes analyzed via an attention mechanism to capture relationships and enrich diagnostic features; (3) Constant data processed through Kolmogorov–Arnold Networks (KAN). This network efficiently leverages constant feature information to extract high-level features via a multi-layer neural network. Subsequently, the hidden features from these three components are concatenated to form an integrated feature vector. This vector is further processed through a final KAN network to predict the patient’s sepsis mortality risk.

\begin{figure}[htb]
    \centering
    \includegraphics[width=1\textwidth, page=1]{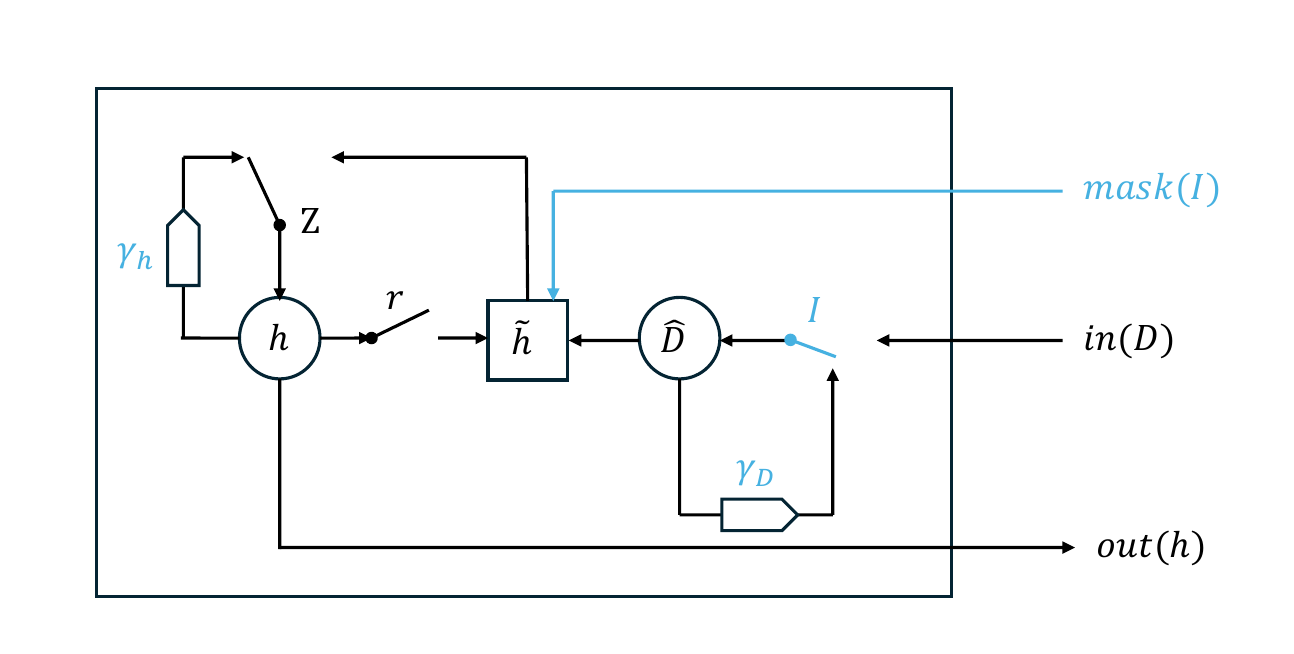}
    \caption{The gated recurrent unit with decay (GRU-D) mechanism, highlighting the interactions between input data, hidden states, and decay factors. The symbols \( \gamma \), \( h \), \( Z \), and \( r \) represent decay factors, hidden state updates, update gates, and reset gates, respectively.}
    \label{fig:figure 4}
\end{figure}

In our methodology, temporal data \(D\) is defined as 
\begin{align*}
D &= (D_1, D_2, \dots, D_t, \dots, D_T) \in \mathbb{R}^{T \times N}
\end{align*}
where \(T\) denotes a 24-hour time interval, and \(N\) represents the number of features. To address missing values, we introduce a mask matrix \(I\) and time intervals \(\Delta\), facilitating effective management of these gaps. The mask \(I \in \mathbb{R}^{T \times N}\) indicates data presence, where a value of 0 shows absence, and 1 shows presence of data at a time step. The time interval \(\Delta_t^n \in \mathbb{R}\) measures the time elapsed since the last observed data point, calculated by:
\[
\Delta_t^n = 
\begin{cases} 
1 + \Delta_{t-1}^n, & \text{if } t > 1 \text{ and } I_{t-1}^n = 0, \\
1, & \text{if } t > 1 \text{ and } I_{t-1}^n = 1, \\
0, & \text{if } t = 1.
\end{cases}
\]
This data is processed through the GRU-D (Gated Recurrent Unit with Decay) model. The GRU-D network effectively manages missing values by utilizing the mask \(I\) and time intervals \(\Delta\). The decay mechanism dynamically adjusts imputation values based on these intervals, thereby mitigating the impact of missing data during state updates. Through this approach, the GRU-D model generates reliable hidden feature representations \(h_D\). Figure ~\ref{fig:figure 4} illustrates the GRU-D network, detailing the operations as follows:

\textbf{Decay factor calculation:}
\[
\gamma_t = \exp\left(-\max(0, W_{\Delta} \Delta_t + b_{\Delta})\right)
\]

\textbf{Input gate:}
\[
Z_t = \sigma\left(W_z (D_t \odot I_t) + U_z (\gamma_t \odot h_{t-1}) + b_z\right)
\]

\textbf{Reset gate:}
\[
r_t = \sigma\left(W_r (D_t \odot I_t) + U_r (\gamma_t \odot h_{t-1}) + b_r\right)
\]

\textbf{Candidate hidden state:}
\[
\tilde{h}_t = \tanh\left(W_h (D_t \odot I_t) + U_h (r_t \odot (\gamma_t \odot h_{t-1})) + b_h\right)
\]

\textbf{Hidden state update:}
\[
h_t = (1 - Z_t) \odot h_{t-1} + Z_t \odot \tilde{h}_t
\]

\textbf{Imputation of missing values:}
\[
\hat{D}_t^n = I_t^n D_t^n - (1 - I_t^n) (\gamma_{D_t}^n D_{t'}^n + (1 - \gamma_{D_t}^n) \tilde{D}^n)
\]
where  \( D_t \) is the input feature vector at time step \( t \), \( I \) is the mask matrix, \( \Delta \) is the time interval matrix, \( \gamma_t \) is the decay factor dynamically adjusted based on the time interval \( \Delta_t \), \( \sigma \) is the sigmoid activation function, \( W \), \( U \), and \( b \) are the weight matrices and bias terms.

In the analysis of the patient's diagnostic ICD code sequence and the corresponding medical ontology CCS sequence, we employ an attention mechanism to elucidate the complex relationships and semantic nuances between the codes. This mechanism calculates the similarity between codes to determine their relative importance and assigns attention weights accordingly, thereby accentuating crucial diagnostic features. We embed the ICD and CCS code sequences, as detailed in section 2.2, into a low-dimensional vector space. Subsequently, we apply the attention mechanism using the following equations:
\[
h_{\text{icd}} = \text{ScaleDotProductAttention}(Q, K, V) = \text{softmax}\left(\frac{Q K^T}{\sqrt{d^k}}\right) V
\]
where \(X_{\text{icd}}\) serves as the query, while \(X_{\text{ccs}}\) functions as both the key and value. The terms \(X_{\text{icd}}\) and \(X_{\text{ccs}}\) denote the ICD and CCS code features, respectively, which are extracted using the Graph Convolutional Network (GCN) as outlined in section 2.2.

\begin{figure}[htb]
    \centering
    \includegraphics[width=1\textwidth, page=1]{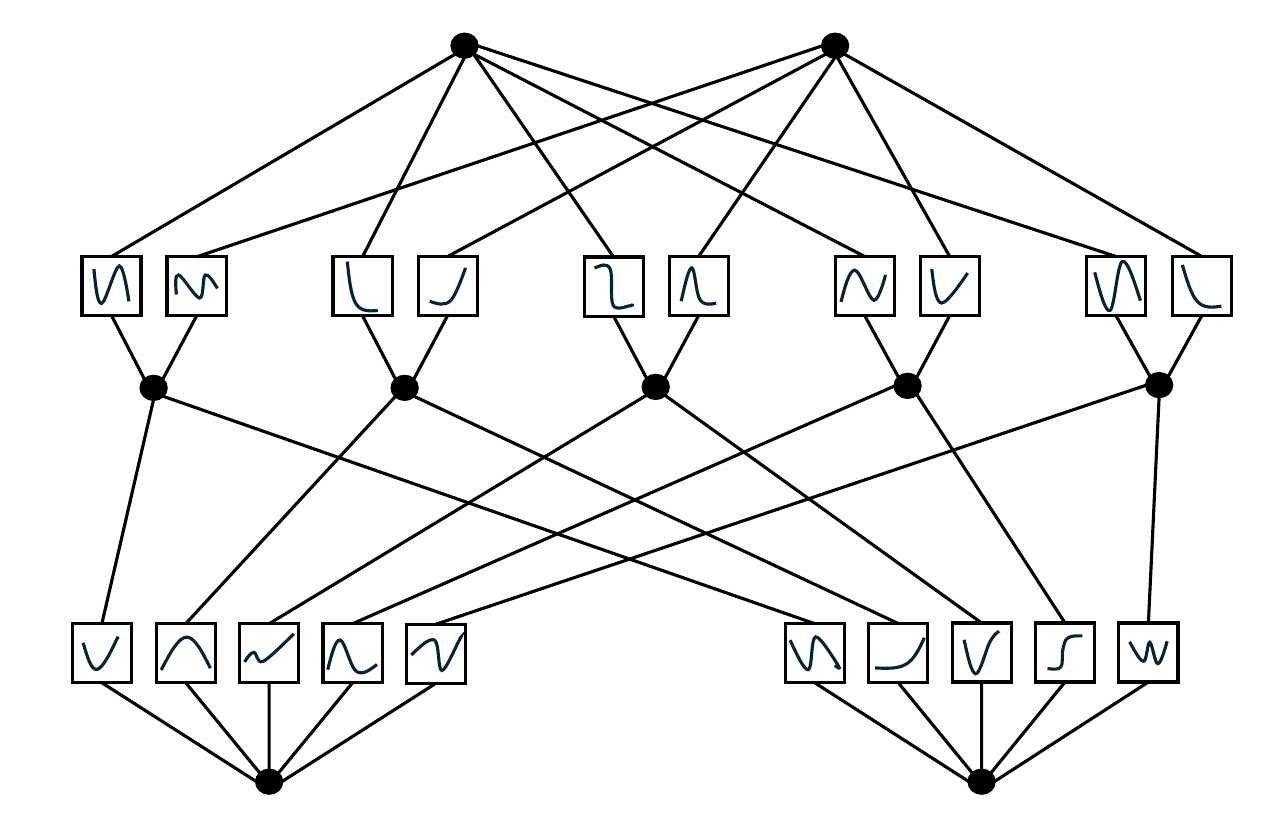}
    \caption{The architectural layout of the Kolmogorov–Arnold Network (KAN), the network's unique structure that employs learnable B-spline activation functions.}
    \label{fig:figure 5}
\end{figure}

We process constant data using the Kolmogorov–Arnold Networks (KAN), which includes basic demographic characteristics of patients such as gender, age, and weight. The KAN network's unique architecture offers greater accuracy and interpretability compared to traditional multilayer perceptrons (MLPs) when handling high-dimensional, multivariate constant data. This network effectively approximates complex functional relationships by employing learnable B-spline activation functions at the edges, enhancing flexibility and adaptability to different data features without significant computational overhead. Unlike traditional MLPs that use fixed activation functions, KAN's design enables precise approximation of multivariate functions and adaptability through learnable activation functions positioned on each layer's edges. Figure~\ref{fig:figure 5} illustrates the KAN network architecture. Each layer in KAN is expressed as a function matrix: \(\Phi = \{\phi_{q,p}\}, p = 1, 2, \ldots, n_{\text{in}}, q = 1, 2, \ldots, n_{\text{out}}\), where \(\phi_{q,p}\) includes trainable parameters. The operational flow within the network is as follows:

\textbf{Pre-activation:} 
\[
x_{l,j,i} = \phi_{l,j,i}(x_{l,i})
\]
where \(x_{l,j,i}\) denotes the pre-activation value from input node \(i\) to output node \(j\) in layer \(l\).

\textbf{Post-activation:}
\[
\tilde{x}_{l,j,i} = \phi_{l,j,i}(x_{l,i})
\]
where \(\tilde{x}_{l,j,i}\) represents the post-activation value from input node \(i\) to output node \(j\) in layer \(l\), after being processed by the function \(\phi_{l,j,i}\).

\textbf{Node Activation:} 
The activation value at node \((l+1,j)\) is the sum of all input activation values:
\[
x_{l+1,j} = \sum_{i=1}^{n_l} \tilde{x}_{l,j,i} = \sum_{i=1}^{n_l} \phi_{l,j,i}(x_{l,i})
\]

In matrix form, this is represented as \(x_{l+1,j} = \Phi_l \cdot x_l\), where \(\Phi_l\) is the function matrix of layer \(l\) in the KAN. The KAN network in this paper is a combination of two layers: \(KAN(x) = (\Phi_1 \circ \Phi_0) x\).
Expanding, this can be written as:
\[
h_s = KAN(x) = \sum_{i_1 = 1}^{n_1} \left( \phi_{1, i_2, i_1} \left( \sum_{i_0 = 1}^{n_0} \phi_{0, i_1, i_0} (x_{i_0}) \right) \right)
\]
where \(n_1\) and \(n_0\) represent the number of nodes in each layer. Finally, the constant data produces hidden features \(h_s\) through the KAN network.

Finally, we concatenate the temporal features \(h_D\), constant features \(h_s\), and diagnostic features \(h_{\text{icd}}\) to create an integrated feature vector. This vector is then used to generate the patient's sepsis mortality risk prediction via a single-layer Kolmogorov–Arnold Network (KAN).

\subsection{Implementation}
In our research, we implemented and trained our model using the TensorFlow framework. During the training process, we systematically optimized several key parameters to improve the model's performance including learning rate, batch size and number of iterations. We utilized the Adam optimizer to update model parameters and employed a learning rate decay strategy to dynamically adjust the learning rate, enhancing the model's convergence stability. To address the significant imbalance between positive and negative samples, we applied oversampling techniques to maintain a consistent ratio of these samples across training, validation, and test sets. This method effectively mitigated model bias toward the majority class, thereby enhancing its generalization capabilities on imbalanced datasets. Moreover, we adopted five-fold cross-validation to validate the model's stability and generalizability. By dividing the dataset into five mutually exclusive subsets, using one for validation and the remaining four for training—and repeating this process five times—we ensured the model's consistent performance across various data splits. To prevent overfitting, we incorporated Dropout layers and monitored the validation set performance using early stopping techniques, halting training when no further reduction in validation loss was observed.

\section{Results}

\subsection{Performance Metrics}
In this section, we describe the metrics employed to assess the TCKAN model's performance. These metrics are vital for gauging the model's accuracy in predicting mortality risk among ICU sepsis patients. The principal evaluation metrics include sensitivity, specificity, area under the receiver operating characteristic curve (AUROC), area under the precision-recall curve (AUPRC), and the Brier Score (BS). Each metric offers a unique perspective on the model’s performance, enabling a thorough analysis. Sensitivity, also known as recall, measures the proportion of positive samples correctly identified by the model. Specificity assesses the proportion of negative samples accurately recognized by the model. The AUROC evaluates the model's overall classification ability across different thresholds; a value closer to 1 indicates stronger discriminative power. AUPRC is a metric that balances precision and recall, particularly useful in cases of imbalanced class distribution. The Brier Score gauges the accuracy of probabilistic predictions, with lower values signifying better accuracy. The specific formulas for these metrics are as follows:

\[
\text{Sensitivity} = \frac{TP}{TP + FN}
\]
\[
\text{Specificity} = \frac{TN}{TN + FP}
\]
\[
\text{Precision} = \frac{TP}{TP + FP}
\]
\[
\text{BS} = \frac{1}{N} \sum_{i=1}^N (f_i - o_i)^2
\]
where \(TP\) is true positive, \(FN\) is false negative, \(TN\) is true negative, \(FP\) is false positive, \(N\) is the number of samples, \(f_i\) is the predicted probability for the \(i\)th sample, and \(o_i\) is the actual label (0 or 1) for the \(i\)th sample.

\subsection{Model Performance}
\begin{figure}[htb]
    \centering
    \includegraphics[width=1\textwidth]{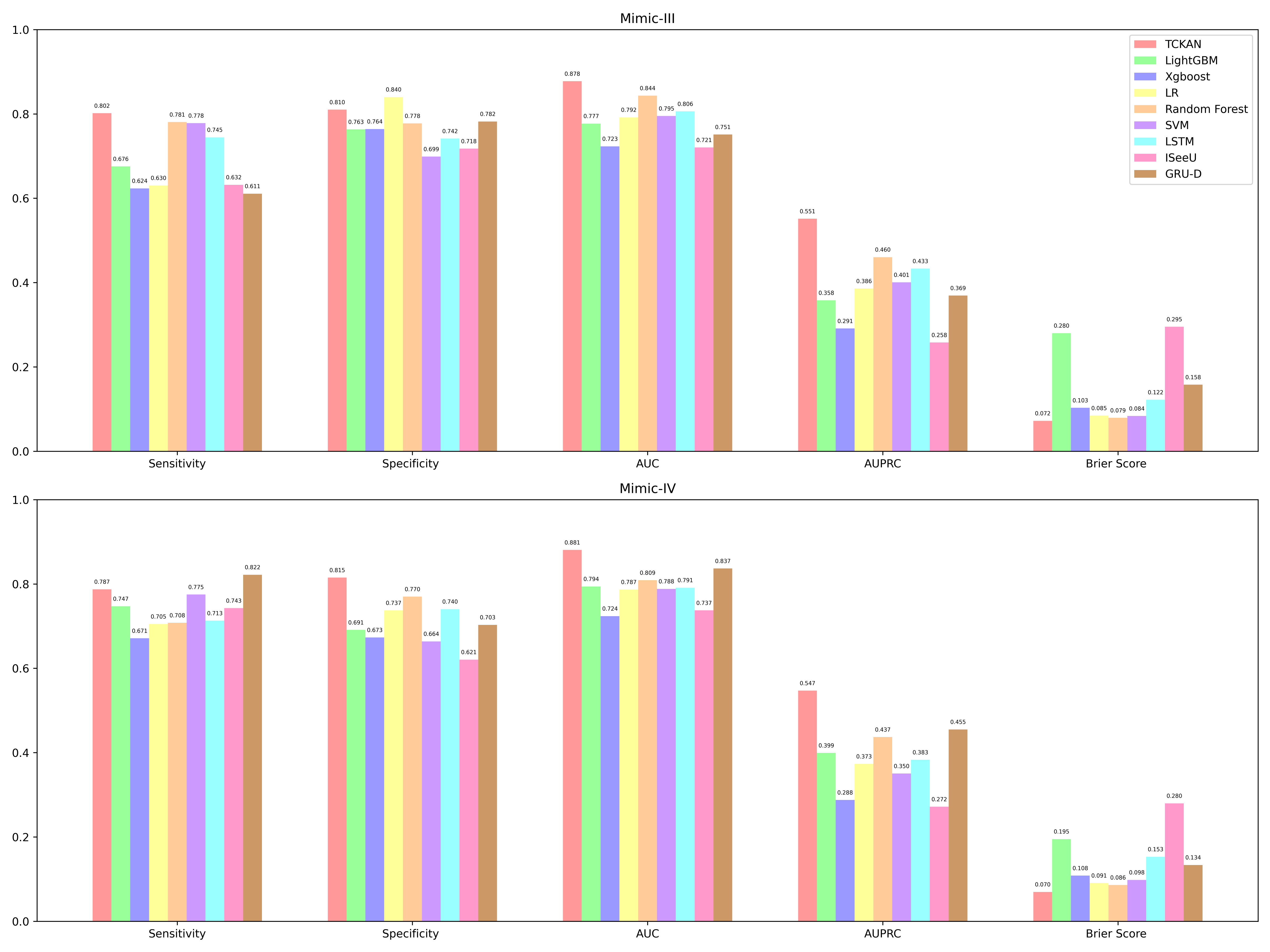}
    \caption{Comparison of performance metrics across various models for predicting mortality risk in ICU sepsis patients using MIMIC-III and MIMIC-IV datasets.}
    \label{fig:figure 6}
\end{figure}

\begin{figure}[htb]
    \centering
    \includegraphics[width=\textwidth]{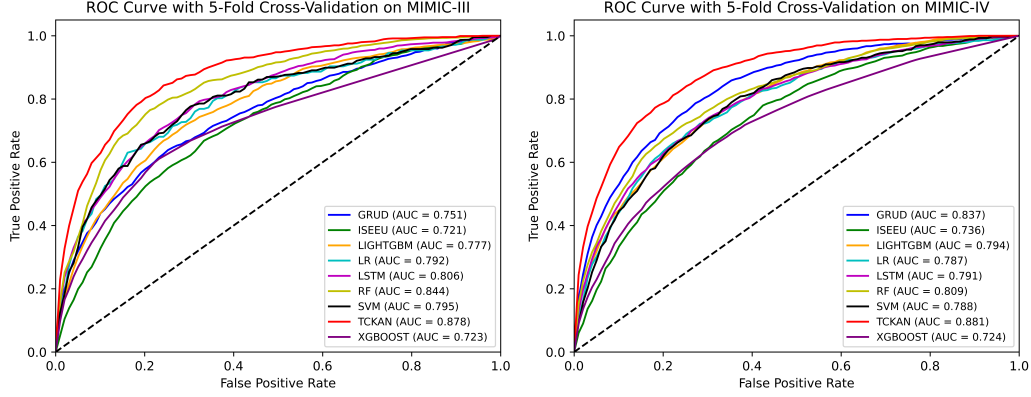}
    \caption{Receiver Operating Characteristic (ROC) curves, employing 5-fold cross-validation on MIMIC-III and MIMIC-IV datasets, display the performance of various models. These include GRUD, ISEEU, LightGBM, Logistic Regression (LR), Long Short-Term Memory (LSTM), Random Forest (RF), Support Vector Machine (SVM), TCKAN, and XGBoost.}
    \label{fig:figure 7}
\end{figure}

We evaluated the TCKAN model's performance on the MIMIC-III and MIMIC-IV datasets by comparing it with seven established baseline models: Xgboost\cite{38}, SVM, Random Forest (RF)\cite{39}, LGBM\cite{22}, LSTM\cite{31}, IseeU\cite{40}, and and Che et al.'s model GRU-D\cite{33}. This comparison across various metrics such as sensitivity, specificity, AUROC, AUPRC, and Brier score allowed us to thoroughly assess the strengths and weaknesses of each model. Figure~\ref{fig:figure 6} illustrates the comparative performance, demonstrating that the TCKAN model generally outperforms the baseline models. Notably, on the MIMIC-IV dataset, the TCKAN model achieved AUROC and AUPRC values of 0.8807 and 0.5470, respectively, significantly surpassing its competitors. This indicates its superior accuracy and stability in distinguishing between positive and negative cases and handling imbalanced datasets. Additionally, the TCKAN model's balanced performance in sensitivity (0.7875) and specificity (0.8153) underscores its effectiveness in identifying patients at different levels of risk. A lower Brier score of 0.0696 reflects the model’s precision in probability predictions. Figure~\ref{fig:figure 7}, showing the ROC curves, further highlights the TCKAN model’s enhanced ability to differentiate between outcomes, with its curve closer to the top-left corner, symbolizing higher prediction accuracy across various thresholds.

\begin{figure}[htb]
    \centering
    \includegraphics[width=1\textwidth]{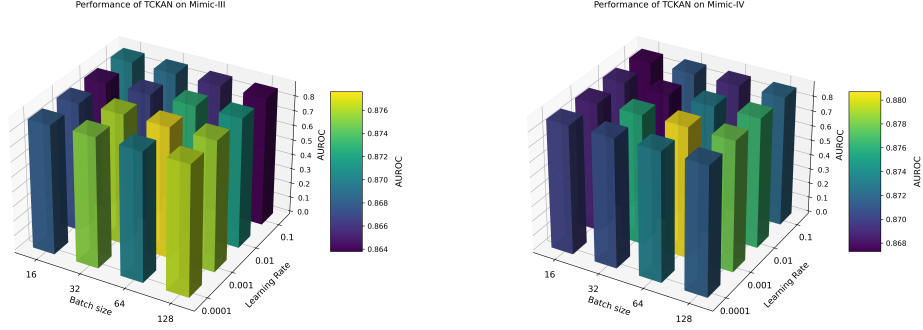}
    \caption{The impact of varying batch sizes and learning rates on the performance of the TCKAN model across MIMIC-III and MIMIC-IV datasets. Each bar represents the Area Under the Receiver Operating Characteristic Curve (AUROC), with colors indicating different AUROC values as batch size and learning rate change.}
    \label{fig:figure 8}
\end{figure}

To further evaluate the robustness of the TCKAN model, we conducted parameter sensitivity experiments focusing on the learning rate (LR) and batch size, two crucial hyperparameters that influence model performance. We systematically varied these parameters, setting the learning rate at 0.0001, 0.001, 0.01, and 0.1, and the batch size at 16, 32, 64, and 128, to observe their effects on model performance under various combinations. Figure~\ref{fig:figure 8} illustrates these performance changes through a three-dimensional graph that plots AUROC values against variations in learning rate and batch size. The results indicate that the TCKAN model achieves optimal performance with a learning rate of approximately 0.001 and a batch size of 64. Notably, excessively high or low learning rates can degrade performance, while very small batch sizes can introduce instability during training, and overly large batch sizes may reduce training speed and impair generalization.

\subsection{Ablation Study}
To further validate the effectiveness and importance of the components of the TCKAN model, we conducted ablation experiments. These experiments involved removing or replacing key modules to assess their impact on the overall model performance. In the first ablation study, we substituted the GRU-D (Gated Recurrent Unit with Decay) module, which is tailored to manage irregular time series data more effectively, with a standard GRU module. This change allowed us to directly compare the performance enhancements offered by GRU-D. Results, as shown in Table \ref{tab:Table 4}, indicate a decline in the model’s AUROC value from 0.8807 to 0.8547 and the AUPRC value from 0.5470 to 0.4957, alongside decreases in sensitivity and specificity. This underscores the critical role of GRU-D in bolstering the accuracy and stability of the predictions.

In the second ablation study, we replaced the two KAN (Kolmogorov-Arnold Network) modules, known for their use of learnable rather than fixed activation functions based on the Kolmogorov-Arnold representation theorem, with multilayer perceptrons (MLP). This modification aimed to evaluate KAN's contribution to the model’s performance. The findings presented in Table \ref{tab:Table 4} reveal that substituting KAN with MLP led to a reduction in the AUROC value from 0.8807 to 0.8693 and the AUPRC value from 0.5470 to 0.5304, with corresponding reductions in sensitivity and specificity. These results highlight the essential role of KAN modules in enhancing the model's capability to discern complex features.

These ablation studies distinctly demonstrate the significant benefits of the GRU-D module in managing irregular time series data and the vital contribution of the KAN module in augmenting the model’s feature recognition capabilities through learnable activation functions.

\begin{table}[]
\centering
\caption{Performance Comparison of TCKAN and Its Variants on MIMIC-III and MIMIC-IV Datasets (mean ± std). All results were validated using 5-fold cross-validation.}
\label{tab:Table 4}
\resizebox{\textwidth}{!}{%
\begin{tabular}{@{}ccccccc@{}}
\toprule
\textbf{Dataset} & \textbf{Model} & \textbf{Specificity} & \textbf{Sensitivity} & \textbf{AUC} & \textbf{Brier Score} & \textbf{AUPRC} \\ \midrule
\multirow{3}{*}{\textbf{MIMIC-III}} & TCKAN                                                            & 0.8017 $\pm$ 0.0169  & 0.8102 $\pm$ 0.0305 & 0.8776 $\pm$ 0.0063 & 0.0721 $\pm$ 0.0024 & 0.5515 $\pm$ 0.0227 \\
                                    & \begin{tabular}[c]{@{}c@{}}TCKAN without GRU-D\end{tabular} & 0.7891 $\pm$ 0.0454 & 0.7681 $\pm$ 0.0390 & 0.8519 $\pm$ 0.0070 & 0.0779 $\pm$ 0.0021 & 0.4803 $\pm$ 0.0187 \\
                                    & \begin{tabular}[c]{@{}c@{}}TCKAN without KAN\end{tabular}    & 0.8092 $\pm$ 0.0264 & 0.7607 $\pm$ 0.0311 & 0.8587 $\pm$ 0.0120 & 0.0738 $\pm$ 0.0019 & 0.5211 $\pm$ 0.0225 \\ \midrule 
\multirow{3}{*}{\textbf{MIMIC-IV}}  & TCKAN                                                            & 0.7875 $\pm$ 0.0298 & 0.8153 $\pm$ 0.0318 & 0.8807 $\pm$ 0.0026 & 0.0696 $\pm$ 0.0006 & 0.5470 $\pm$ 0.0088 \\
                                    & \begin{tabular}[c]{@{}c@{}}TCKAN without GRU-D\end{tabular}  & 0.8063 $\pm$ 0.0359 & 0.7635 $\pm$ 0.0259 & 0.8547 $\pm$ 0.0123 & 0.0748 $\pm$ 0.0013 & 0.4957 $\pm$ 0.0259 \\
                                    & \begin{tabular}[c]{@{}c@{}}TCKAN without KAN\end{tabular}    & 0.8173 $\pm$ 0.0202 & 0.7787 $\pm$ 0.0114 & 0.8693 $\pm$ 0.0064 & 0.0716 $\pm$ 0.0007 & 0.5304 $\pm$ 0.0081 \\ \midrule 
\end{tabular}%
}
\end{table}

\section{Discussion}
This study analyzed key data features influencing sepsis mortality prediction, revealing several critical factors that enhance predictive accuracy. Previous research has demonstrated the importance of accurate and timely detection models in various medical emergencies. Similarly, these methodologies underscore the potential of machine learning in enhancing sepsis mortality prediction by integrating multiple data types.

\begin{figure}[htb]
    \centering
    \includegraphics[width=1\textwidth]{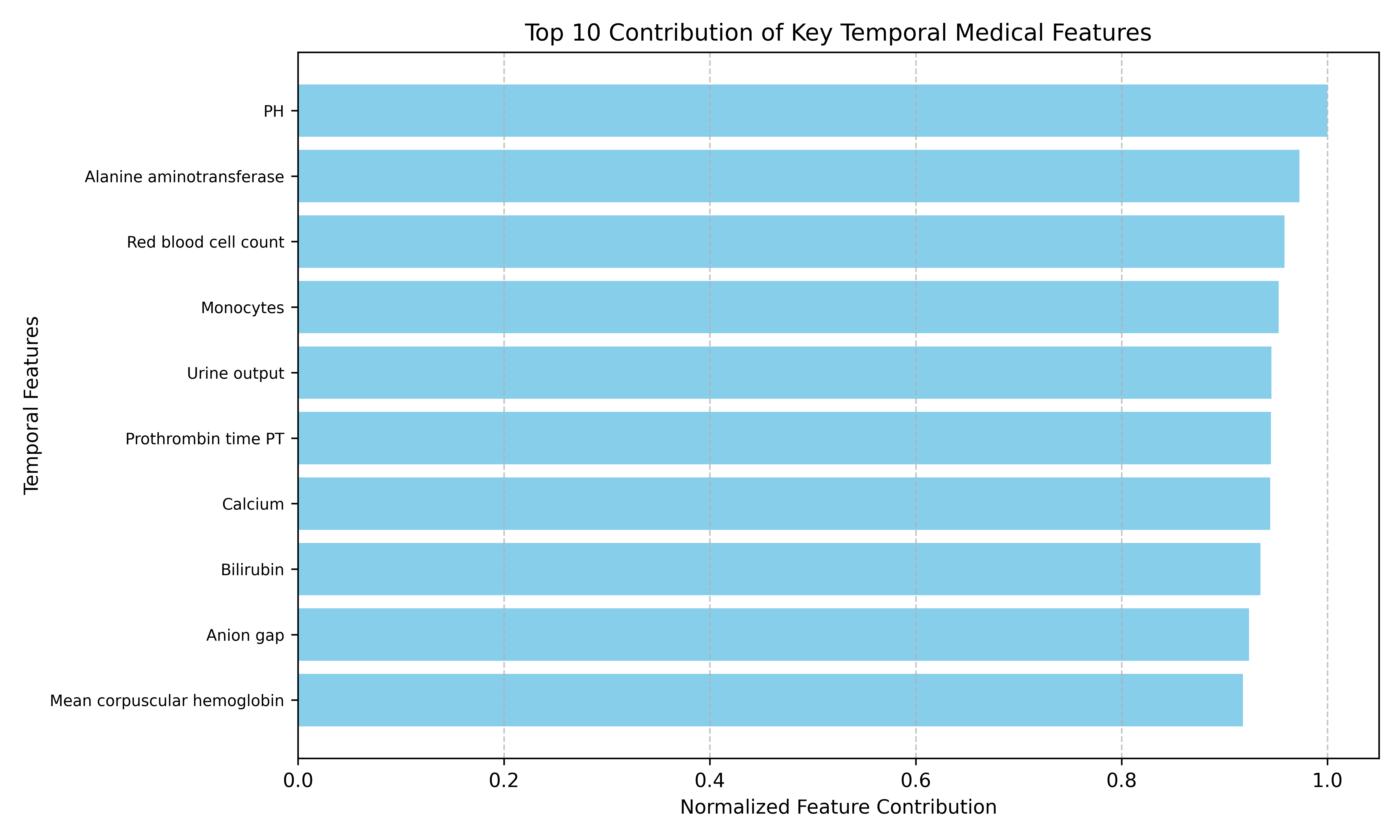}
    \caption{The normalized contribution of the top 10 key temporal medical features in TCKAN model.}
    \label{fig:figure 9}
\end{figure}

For the temporal data, Figure~\ref{fig:figure 9} highlights the top ten features with the highest feature weights and their contributions, including pH value, alanine aminotransferase, red blood cell count, monocyte count, urine output, prothrombin time, calcium levels, bilirubin, anion gap, and mean corpuscular hemoglobin concentration. These indicators not only reflect the patient’s current physiological state but also detect subtle shifts in their condition, enhancing the accuracy of mortality predictions. For instance, variations in pH value can signal shifts in acid-base balance, alterations in urine output may indicate changes in renal function, and levels of alanine aminotransferase and bilirubin provide insights into liver health. Additional data such as red blood cell and monocyte counts, along with mean corpuscular hemoglobin concentration, are crucial for assessing blood health and immune response, while prothrombin time and anion gap offer insights into coagulation and metabolic status.

In addition to the previously discussed temporal features, it is essential to consider the effects of sepsis on cardiovascular parameters, particularly heart rate and blood pressure. Sepsis often triggers tachycardia as a compensatory response to systemic vasodilation and hypotension. Blood pressure fluctuations during sepsis are frequently characterized by episodes of hypotension, reflecting the body's struggle to maintain adequate perfusion. Moreover, the strain on the cardiovascular system can result in arrhythmias, which serve as critical indicators of deteriorating cardiac function and are associated with increased mortality rates in septic patients. Future research should integrate these cardiovascular metrics into temporal analyses to provide a more comprehensive understanding of sepsis progression and its impact on patient outcomes.

\begin{figure}[htb]
    \centering
    \includegraphics[width=1\textwidth]{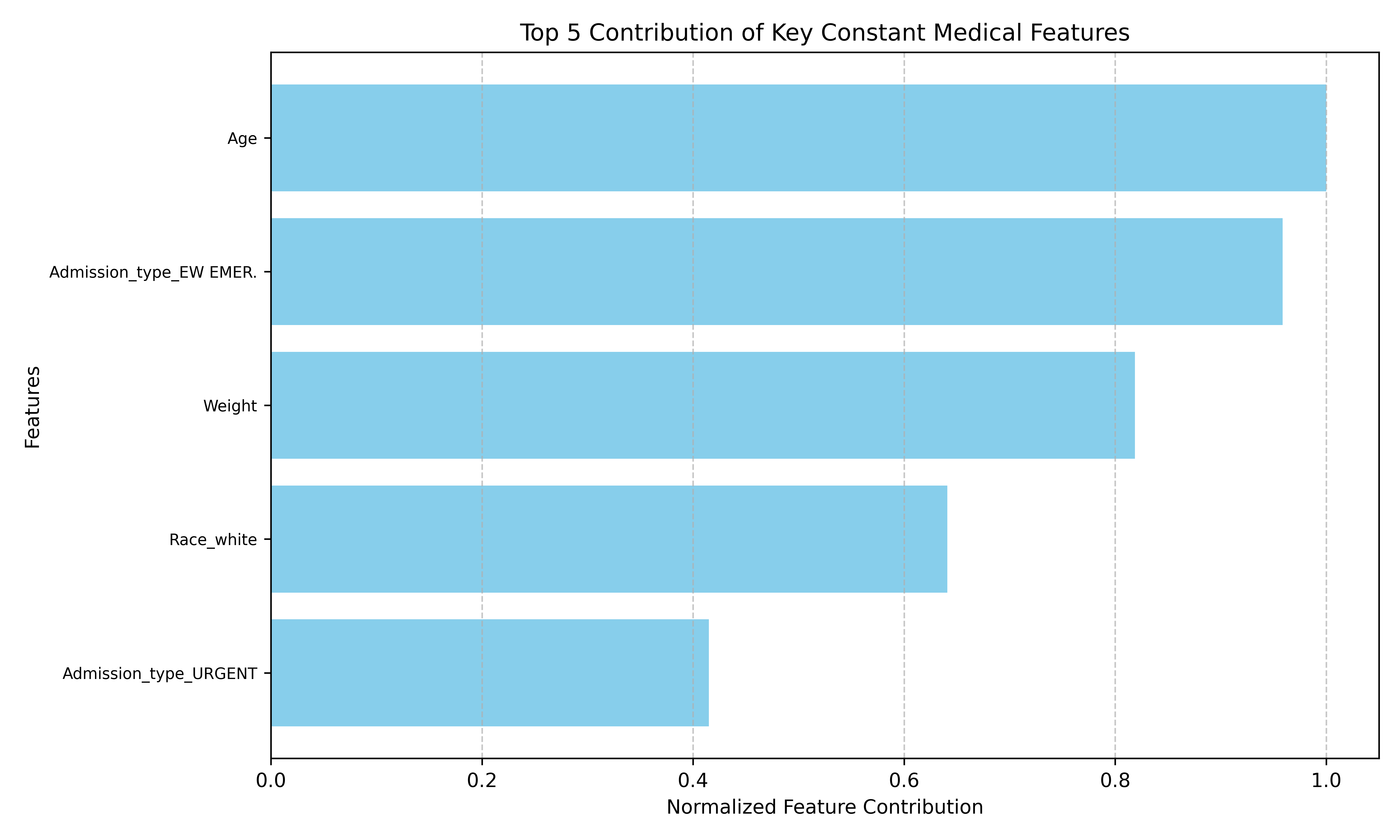}
    \caption{The normalized contribution of the top 5 key constant medical features in TCKAN model.}
    \label{fig:figure 10}
\end{figure}

Figure~\ref{fig:figure 10} presents the top five features by feature weight within the constant data, including age, race, weight, and type of admission. These attributes provide essential health and demographic information about the patient. Notably, age is a crucial factor because older patients often have a higher risk of mortality. Additionally, weight serves as a significant health indicator that can affect the severity of diseases and the efficacy of treatments, thereby influencing the overall prognosis. The nature of admission (emergency, urgent, or elective) informs the urgency of the patient’s condition upon hospitalization.

Our analysis of patient ICD codes revealed that certain codes, especially those related to severe conditions like diabetes and specific malignancies, are strongly associated with increased mortality risk. Codes for secondary diabetes, certain skin disorders, diseases of the middle ear and mastoid, and malignant neoplasms of the retina, for instance, highlight underlying severe health issues that can complicate the patient's recovery process. Additionally, ICD codes for benign tumors of the colon, malignant neoplasms of the cervix, spirochaetal jaundice, neuroendocrine tumors, trigeminal neuralgia, and Pneumocystis jirovecii pneumonia were also identified as significantly influencing mortality rates, underscoring their importance in patient management and treatment focus.

Despite the achievements of this study, it has several limitations. First, the datasets used, primarily MIMIC-III and MIMIC-IV, originate from a single medical center and may not adequately represent patient conditions in other regions or healthcare facilities. Second, while the model's predictive performance surpasses existing methods, opportunities for enhancement remain, particularly in processing specific types of patient data. For instance, certain features may exhibit variable performance across different populations, necessitating further validation of the model's applicability. Additionally, the feature selection and preprocessing methods employed may not capture all potentially significant features, potentially impacting the accuracy of the model’s predictions.

Future research should consider including a broader array of features, such as genetic markers, imaging data, and socioeconomic factors, to provide a more comprehensive assessment of a patient's health condition and to improve the model's predictive capabilities. Additionally, integrating advanced imaging techniques, such as spectral CT material decomposition, can significantly reduce noise and enhance data quality, which in turn may further improve the accuracy of sepsis prediction models \cite{hohweiller2017spectral}. Further validation and optimization could be pursued through comparative studies with other independent datasets, ensuring the model's consistency and reliability across various settings. Through ongoing refinement and expansion, this research is poised to offer more robust support for the early detection and management of sepsis.

\section{Conclusion}
The Temporal-Constant Kolmogorov-Arnold Network (TCKAN) model presented in this study introduces several novel contributions to the field of sepsis mortality prediction. By uniquely integrating constant and temporal data with patient diagnostic ICD codes and medical ontology CCS codes, TCKAN addresses critical limitations of existing models that typically focus on a single data modality. This innovative multi-modal approach not only enhances predictive accuracy but also offers a more comprehensive and nuanced understanding of patient risk profiles, thereby setting a new standard for predictive models in critical care settings. The model outperformed existing methods across several key metrics, including AUROC, AUPRC, sensitivity, and specificity, demonstrating its robustness in handling complex medical data. The incorporation of the GRU-D and KAN modules was particularly beneficial, enhancing the model’s ability to process irregular temporal data and capture intricate features effectively.

The TCKAN model can be deployed in hospitals and Intensive Care Units (ICUs) to monitor and predict the mortality risk of sepsis patients in real-time. By providing accurate risk assessments, it also optimizes the allocation of medical resources, including the distribution of ICU beds, medical staff, and emergency equipment. Moreover, by integrating multi-source data, the model offers more detailed analyses of patient conditions, which aids in the development of personalized treatment plans and improves both treatment outcomes and patient prognosis. Additionally, the TCKAN model can function as a core component of Clinical Decision Support Systems (CDSS), helping doctors make more informed clinical decisions in complex medical environments, thereby enhancing the overall quality of healthcare services.

However, while the TCKAN model has demonstrated impressive predictive accuracy, there remains considerable scope for future research. Future studies could focus on integrating additional types of medical data, such as genetic markers, imaging data, and socioeconomic factors, to provide a more comprehensive assessment of patient health conditions. Additionally, validating the model across diverse datasets from different medical centers will be crucial to ensure its generalizability and applicability in various clinical settings. By continuously refining and expanding the TCKAN model, its utility in early detection and management of sepsis could be further enhanced, potentially leading to better patient outcomes and reduced mortality rates

\section*{Acknowledgments}
This work is supported by the Yunnan Provincial Foundation for Leaders of Disciplines in Science and Technology, China under Grant 202305AC160014.

\section*{Conflict of Interest Statement}
The authors declare that there are no conflicts of interest regarding the publication of this paper.

\bibliography{references}

\end{document}